\documentclass[12pt]{iopart}

\usepackage{epsf,iopams,graphicx}

\newcommand{\lla}{\left\langle}
\newcommand{\rra}{\right\rangle}

\begin{document}

\title
{Recovery of the fidelity amplitude for the Gaussian ensembles}

\author{H-J St\"ockmann, R Sch\"afer}



\address{Fachbereich Physik der Philipps-Universit\"at Marburg,
D-35032 Marburg, Germany}


\begin{abstract}
Using supersymmetry techniques analytical expressions for the average of the fidelity
amplitude $f_{\epsilon}(\tau)=\left<\psi(0)|\exp(2\pi\imath H_{\epsilon}\tau)
\exp(-2\pi\imath H_{0}\tau)|\psi(0)\right>$ are obtained, where $
H_\epsilon=H_0+(\sqrt{\epsilon}/2\pi) V$, and $H_0$ and $H_{\epsilon}$ are taken from the
Gaussian unitary ensemble (GUE) or the Gaussian orthogonal ensemble (GOE), respectively.
As long as the perturbation strength is small compared to the mean level spacing, a
Gaussian decay of the fidelity amplitude is observed, whereas for stronger perturbations
a change to a single-exponential decay takes place, in accordance with results from
literature. Close to the Heisenberg time $ \tau=1$, however, a partial revival of the
fidelity is found, which hitherto remained unnoticed. Random matrix simulations have been
performed for the three Gaussian ensembles. For the case of the GOE and the GUE they are
in perfect agreement with the analytical results.
\end{abstract}



\ead{stoeckmann@physik.uni-marburg.de}

\maketitle











\newcommand {\xb} {{\bf x}}
\newcommand {\xbd} {{\bf x^\dag}}
\newcommand {\yb} {{\bf y}}
\newcommand {\ybd} {{\bf y^\dag}}
\newcommand {\zn} {{\bf z}_n}
\newcommand {\znd} {{\bf z}_n^{\bf\dag}}
\newcommand {\ree} {R_\epsilon\left(E_1,E_2\right)}

\newcommand {\bet}{B}
\newcommand {\gam}{C}

\newcommand {\diag} {{\rm diag}}

\section{Introduction}

The concept of fidelity has been developed as a tool to characterize the stability of a
quantum-mechanical system against perturbations \cite{per84}. Originally fidelity was
introduced as the squared modulus of the overlap integral of a wave packet with itself
after the development forth and back under the influence of two slightly different
Hamiltonians. Let $H_0$ be the unperturbed Hamiltonian and
\begin{equation}\label{000}
    H_\epsilon=H_0+\frac{\sqrt{\epsilon}}{2\pi} V
\end{equation}
the perturbed one. This somewhat unusual definition of the perturbation strength
$\epsilon$ has been applied for later convenience. Then the fidelity is given by
\begin{equation}\label{00}
  F_{\epsilon}(\tau)=\left|\left<\psi(0)|\exp(2\pi\imath
  H_\epsilon\tau)
\exp(-2\pi\imath H_{0}\tau)|\psi(0)\right>\right|^2\,,
\end{equation}
where $\psi(0)$ is the wave function at the beginning, often chosen as a Gaussian wave
packet with minimum uncertainty. It is assumed that $H_0$ has mean level spacing of one,
and thus $\tau$ is given in units of the Heisenberg time. The variance of the
off-diagonal elements of $V$ is chosen to be one. In all what follows it is assumed that
$\epsilon$ is of the order of one thus guaranteeing that the shift of the levels due to
the parameter variation is of the order of the mean level spacing.

Depending on the strength of the perturbation one can discriminate roughly three regimes.
In the perturbative regime, where the strength of the perturbation is small compared to
the mean level spacing, the decay of the fidelity is Gaussian. As soon as the strength of
the perturbation becomes of the order of the mean level spacing, a cross-over to
exponential decay is observed, with a decay constant obtained from Fermi's golden rule
\cite{cer02,jac01b}. For very strong perturbations the decay becomes independent of the
strength of the perturbation. Here the decay is still exponential, but now the decay
constant is given by the classical Lyapunov exponent \cite{Jal01}.   It has been
proposed by Pastawski \etal \cite{pas95} to look for such a behaviour in a spin-echo
experiment on isolated spins coupled weakly to a bath of surrounding spins \cite{zha92}.

A paper of Gorin \etal \cite{gor04} is of particular relevance for the present work. The
authors calculated the Gaussian average of the fidelity amplitude in the regime of small
perturbations using the linear-response approximation,
\begin{equation}\label{00a}
  f_{\epsilon}(\tau)\sim 1- \epsilon \, C(\tau)\,.
\end{equation}
where $C(\tau)$ is given by
\begin{equation}\label{00aa}
    C(\tau)=\frac{\tau^2}{\beta}+\frac{\tau}{2}-\int_0^\tau \int_0^t
b_{2,\beta}(t^\prime) \rmd t^\prime \rmd t \, ,
\end{equation}
and $1-b_{2,\beta}(\tau)$ is the spectral form factor. $\beta$ is the universality index, i.\,e. $\beta=1$ for the
Gaussian orthogonal ensemble (GOE), $\beta=2$ for the Gaussian unitary ensemble (GUE),
and $\beta=4$ for the Gaussian symplectic ensemble (GSE). For an explicit calculation
knowledge of the spectral form factor is thus sufficient. By an exponentiation of the
above formula,
\begin{equation}\label{00b}
  f_{\epsilon}(\tau)\sim e^{-\epsilon \, C(\tau)}\,,
\end{equation}
the authors were able to describe the cross-over from Gaussian to exponential decay with
increasing perturbation strength quantitatively. The Lyapunov regime is
non-universal and thus not accessible in a random matrix model.

It is obvious that the linear-response approximation must break down for large perturbations. 
In the present work supersymmetry techniques are applied to calculate the ensemble 
average of the fidelity decay. Since this calculation is non-perturbative, the results hold for
arbitrary values of the perturbation strength. 
We shall see that the calculation reveals an important generic feature, which is unaccessible by any perturbative approach.

\section{The model}

In the present paper we shall discuss the ensemble average of the {\em fidelity
amplitude}, since for this quantity the calculation is much easier than for the
originally introduced quantity (\ref{00}). 
Since both the unperturbed Hamiltonian and the perturbation are taken from the Gaussian ensembles, 
the choice of the initial wave packet $\psi(0)$ is irrelevant.  
The ensemble average may thus be written as
\begin{equation}\label{01}
  f_\epsilon(\tau)=\frac{1}{N}\left<{\rm Tr}\left[e^{2\pi\imath H_\epsilon\tau}
  e^{-2\pi\imath H_0\tau}\right]\right>\,,
\end{equation}
where it is assumed that the Hamiltonian has been truncated to a finite rank $N$.

The Hamiltonian introduced in equation (\ref{000}) has the disadvantage that the mean
density of states changes with $\epsilon$. The more it is somewhat inconvenient for the
present calculation that the variances of the matrix elements of $H_0$ and $V$ differ. We
therefore adopt a slightly different parameter variation,
\begin{equation}\label{00c}
  H_\phi=H_0\cos\phi+ H_1\sin\phi\,.
\end{equation}
It is assumed that  the matrix elements of $H_0$ and $H_1$  have the same variance,
 have zero average, $\langle (H_0)_{ij}\rangle =\langle (H_1)_{ij} \rangle=0$,
are uncorrelated, $\left<(H_0)_{ij}(H_1)_{kl}\right>=0$, and are Gaussian
distributed,
\begin{equation}\label{00e}
  p\left(\left\{(H_n)_{kl}\right\}\right)\sim
  \exp\left(-\frac{1}{2\lambda_\beta}\Tr H_n^2\right)\,,\qquad
  n=0,1\,,
\end{equation}
where
\begin{equation}\label{19}
  \lambda_\beta=\frac{2N}{\beta\pi^2}\,.
\end{equation}

From random matrix theory it is known that the corresponding ensemble averaged density of
states is given by Wigner's semi-circle law with a value of one in the centre of the
circle,
\begin{equation}\label{18}
  \rho(\bar{E})=
  \sqrt{1-
  \left(\frac{\pi\bar{E}}{2N}\right)^2}
\end{equation}
(see e.\,g. reference \cite{meh91}). The resulting variances of the matrix elements of
$H_0$ and $H_1$ are given by
\begin{equation}
\label{18a} \lla | (H_n)_{kl} |^2 \rra = \frac{N}{\pi^2} \left\{\begin{array}{ll}
    1\,,\qquad & k \neq l \\
    \frac{2}{\beta}\,,& k=l
  \end{array}\right. \qquad n=0,1\,.
\end{equation}
It follows for the GUE
\begin{equation}\label{00g}
  \left<[H_{\phi_2}]_{kl}^*[H_{\phi_1}]_{kl}\right>=\frac{N}{\pi^2}
  \cos\left(\phi_2-\phi_1\right)\,.
\end{equation}
A similar expression is obtained for the GOE. Since only the difference of $\phi_2$ and
$\phi_1$ enters expression (\ref{00g}),
 we may assume without loss of generality $\phi_1=-\phi_2=\phi/2$.
Ansatz (\ref{00c}) for the parameter variation is obtained from equation (\ref{000}) by
means of the substitutions
\begin{equation}\label{00f}
    H_0\to\cos\phi\, H_0\,,\qquad
    V\to\frac{\pi}{\sqrt{N}}\cos\phi\,
    H_1\,,
\end{equation}
where
\begin{equation}\label{00h}
    \tan\phi=\sqrt{\frac{\epsilon}{4N}}\,.
\end{equation}
$\phi$ is thus of ${\cal O}(\frac{1}{\sqrt{N}})$, and $\epsilon$ is given in the limit of
large $N$ by
\begin{equation}\label{00i}
    \epsilon=4N\phi^2\,.
\end{equation}
Only terms up to ${\cal O}(\phi^2)$ will survive the limit $N\to\infty$ as we shall see
later. The details of the parameter dependence are irrelevant.

With all these substitutions equation (\ref{01}) may be transformed into
\begin{equation}\label{02}
  f_\epsilon(\tau)=\int dE_1\,dE_2
  e^{2\pi\imath\left(E_1-E_2\right)\tau}\,\ree\,
\end{equation}
where
\begin{equation}\label{03}
  \ree\sim\frac{1}{N}\left<{\Tr}\left(
  \frac{1}{E_{1-}-cH_0-sH_1}\,\frac{1}{E_{2+}-cH_0+sH_1}\right)\right>\,,
\end{equation}
with $E_\pm=E\pm\imath\eta$, and the abbreviations $c=\cos(\phi/2)$, $s=\sin(\phi/2)$.
Using standard supersymmetry techniques \cite{ver85a}, this can be written as
\begin{eqnarray}\label{04}
  \lefteqn{\ree}\nonumber \\&\sim&\frac{1}{N}
  \Bigg<\int d[x]\,d[y]\,\sum\limits_{n,m}
  (x_n^*x_m-\xi_n^*\xi_m)(y_m^*y_n-\eta_m^*\eta_n)\nonumber\\
  &&\times e^{-\imath
  \xbd\left(E_1-cH_0-sH_1\right)\xb}\,
  e^{\imath \ybd\left(E_2-cH_0+sH_1\right)\yb}\Bigg>
  \nonumber \\&=&\frac{1}{N}
  \int d[x]\,d[y]\,\sum\limits_{n,m}
  (x_n^*x_m-\xi_n^*\xi_m)(y_m^*y_n-\eta_m^*\eta_n) e^{-\imath\left[\xbd E_1\xb-\ybd
  E_2\yb\right]}\nonumber\\&&\times
  \left<e^{\imath c\left[\xbd H_0\xb-\ybd H_0 \yb\right]}\right>\,
  \left<e^{\imath s\left[\xbd H_1\xb+\ybd H_1 \yb\right]}\right>\,,
\end{eqnarray}
where $\xb=\left(x_1,\xi_1,\dots,x_N,\xi_N\right)^T$,
$\yb=\left(y_1,\eta_1,\dots,y_N,\eta_N\right)^T$, and

\begin{equation}\label{05}
d[x]=\prod_n dx_n\,dx_n^*\,d\xi_n\,d\xi_n^*\,,\qquad d[y]=\prod_n
dy_n\,dy_n^*\,d\eta_n\,d\eta_n^*\,.\nonumber
\end{equation}
We adopt the usual convention and use latin letters for commuting, and greek ones for
anticommuting variables, respectively. Equation (\ref{04}) is still true for all Gaussian
ensembles, but now we have to discriminate between the GUE and the GOE.

\section{The GUE case}\label{gue}

Using equation (\ref{00e}), the calculation of the Gaussian average over $H_0$ and $H_1$
is elementary. The result for $H_0$ may be expressed as
\begin{equation}\label{06}
  \left<e^{\imath c\left(\xbd H_0\xb-\ybd H_0\yb\right)}\right>=
e^{-\frac{c^2\lambda_\beta}{2}{\Tr}S^2}\,,
\end{equation}
where
\begin{equation}\label{07}
  S=\sum\limits_n\left(\begin{array}{c}
    x_n \\ \xi_n \\
    -y_n \\ -\eta_n\
  \end{array}\right)\left(x^*_n, \xi^*_n,y^*_n, \eta^*_n\right)\,.
\end{equation}

Whenever supermatrices are involved, traces and determinants are to be interpreted as
{\em super} traces and determinants, respectively, in the definition of reference
\cite{ver85a}. In short hand notation equation (\ref{07}) may be written as
\begin{equation}\label{07a}
  S={\bf L}\sum\limits_n\zn\znd\,,
\end{equation}
where
\begin{equation}\label{07b}
  \zn=\left(\begin{array}{c}
    x_n \\ \xi_n \\
    y_n \\ \eta_n\
  \end{array}\right)\,,\qquad {\bf L}=\left(\begin{array}{cc}
    {\bf 1_2} & \cdot \\
    \cdot & -{\bf 1_2}\
  \end{array}\right)\,,
\end{equation}
and {$\bf 1_2$} is the two-dimensional unit matrix. Introducing the notation
\begin{equation}\label{07c}
  S=\left(\begin{array}{cc}
    S_{AA} & S_{AR} \\
    S_{RA} & S_{RR} \
  \end{array}\right)\,,
\end{equation}
where each $S_{ij}$ is a $ 2\times 2$ matrix, and the indices {\em `A', `R'} refer to the
`advanced' and `retarded' components, respectively, the sum entering equation (\ref{04})
may concisely be written as
\begin{equation}\label{07d}
  \sum\limits_{n,m}
  (x_n^*x_m-\xi_n^*\xi_m)(y_m^*y_n-\eta_m^*\eta_n)=
  -{\rm Tr}( S_{AR}\,\sigma\, S_{RA}\, \sigma)\,,
\end{equation}
where
\begin{equation}\label{07e}
  \sigma=\left(\begin{array}{cc}
    1 & \cdot \\
    \cdot & -1\
  \end{array}\right)\,.
\end{equation}
Next, a Hubbard-Stratonovich transformation is applied to equation (\ref{06}),
\begin{equation}\label{08}
   \left<e^{\imath c\left(\xbd H_0\xb-\ybd H_0\yb\right)}\right>=
   \int d[u] e^{-\frac{1}{2\lambda_\beta}\Tr U^2+\imath c\Tr US}\,,
\end{equation}
where $U$ is the supermatrix
\begin{equation}\label{09}
  U=\left(\begin{array}{cc}
    U_{AA} & U_{AR} \\
    U_{RA} & U_{RR} \
  \end{array}\right)
\end{equation}
with the $2\times 2$ components
\begin{equation}\label{10}
  U_{ij}=\left(\begin{array}{cc}
    u_{ij} & \omega^*_{ij} \\
    \omega_{ij} & \bar{u}_{ij} \
  \end{array}\right)\,,\quad i,j=A,R\,.
\end{equation}

For the integrals in equation~(\ref{08}) to be well-defined, the $u_{ij}$ integrations
have to be performed from $-\infty$ to $\infty$, whereas the $\bar{u}_{ij}$ integrations
are from $-\imath\infty$ to $\imath\infty$. (In literature usually an additional factor
of $\imath$ is introduced in the lower right corner of the matrix (\ref{10}) to avoid
integrations along the imaginary axis.)

In the same way we obtain
\begin{equation}\label{11}
   \left<e^{\imath s(\xbd H_1\xb + \ybd H_1\yb)}\right>=
   \int d[v] e^{-\frac{1}{2\lambda_\beta}\Tr V^2+\imath s\Tr VT}\,,
\end{equation}
where
\begin{equation}\label{12}
  T=\sum\limits_n \zn\znd\,,
\end{equation}
and
\begin{equation}\label{12a}
  V=\left(\begin{array}{cc}
    V_{AA} & V_{AR} \\
    V_{RA} & V_{RR} \
  \end{array}\right)\,,
  \end{equation}
with
  \begin{equation}\label{13}
  V_{ij}=\left(\begin{array}{cc}
    v_{ij} & \nu^*_{ij} \\
    \nu_{ij} & \bar{v}_{ij} \
  \end{array}\right)\,,\quad i,j=A,R \,.
\end{equation}
Collecting the results we obtain from equation (\ref{04})

\begin{eqnarray}\label{13a}
  \ree&\sim&\frac{1}{N}\int d[u]\,d[v]e^{-\frac{1}{2\lambda_\beta}\Tr (U^2+V^2)}\nonumber\\
  &&\times\int d[x,y]
  \Tr( S_{AR}\,\sigma\, S_{RA}\, \sigma) e^{-\imath\left[\xbd E_1\xb-\ybd
  E_2\yb\right]}
  e^{\imath\left[c \Tr(US)+s\Tr(VT)\right]}\nonumber\\
  &\sim&\frac{1}{c^2N}\int d[u]\,d[v]e^{-\frac{1}{2\lambda_\beta}\Tr (U^2+V^2)}
  \Tr\left( \frac{\partial}{\partial U_{AR}}\,\sigma\,
  \frac{\partial}{\partial U_{RA}}\sigma\right)\nonumber\\
  &&\times\int d[x,y]
   e^{-\imath\left[\xbd E_1\xb-\ybd
  E_2\yb\right]}
  e^{\imath\left[c \Tr(US)+s\Tr(VT)\right]}\,.
\end{eqnarray}

Now the $x,y$ integrations can be performed resulting in
\begin{eqnarray}\label{14}
  \ree
  &\sim&\frac{1}{c^2N}\int d[u]\,d[v]e^{-\frac{1}{2\lambda_\beta}\Tr (U^2+V^2)}
  \Tr\left( \frac{\partial}{\partial U_{AR}}\,\sigma\,
  \frac{\partial}{\partial U_{RA}}\sigma\right)|M|^{-N}\nonumber\\
 \end{eqnarray}
where
\begin{equation}\label{15}
  |M|=\left|\begin{array}{cc}
    E_1\,{\bf 1_2}-cU_{AA}-sV_{AA} & cU_{AR}-sV_{AR} \\
    -cU_{RA}-sV_{RA} & -E_2\,{\bf 1_2}+cU_{RR}-sV_{RR} \
  \end{array}\right|\,.
\end{equation}
Introducing the notation $E_{1/2}=\bar{E}\pm E/2$, this can be written as
\begin{equation}\label{15a}
  |M|= \left|\bar{E}\,{\bf 1_4}-cU+\left(\frac{E}{2}-sV\right){\bf L}\right|\,,
\end{equation}
where $\bf 1_4$ is the $4\times 4$ unit matrix, and ${\bf L}$ has been given in equation
(\ref{07b}). Substituting
\begin{equation}\label{15b}
  U=\frac{1}{c}\left[\hat{U}+\left(\frac{E}{2}-sV\right){\bf L}\right]\,,
\end{equation}
the $\hat{V}$ integrations can be performed yielding
\begin{eqnarray}\label{16}
  \lefteqn{\ree}\nonumber\\
  &\sim&\frac{1}{N}\int d[u]
  e^{-\frac{1}{2\lambda_\beta}\Tr
  \left[U_{AA}^2+U_{BB}^2+E\left(U_{AA}-U_{RR}\right)+\frac{2}
  {\cos\phi}\left(U_{AR}U_{RA}\right)\right]}
  \nonumber\\&&\times\Tr\left( \frac{\partial}{\partial U_{AR}}\,\sigma\,
  \frac{\partial}{\partial U_{RA}}\sigma\right)|\bar{E}{\bf 1_4}-U|^{-N}\nonumber\\
&\sim&\frac{1}{N\lambda_\beta^2\cos^2\phi}\int d[u]\Tr\left(U_{AR}\,\sigma\,
  U_{RA}\sigma\right)
 \nonumber\\&&\times e^{-\frac{1}{2\lambda_\beta}\Tr
  \left[E\left(U_{AA}-U_{RR}\right)+2\left(\frac{1}
  {\cos\phi}-1\right)\left(U_{AR}U_{RA}\right)\right]}
  e^{-\Tr g(U)}\,,
\end{eqnarray}
where
\begin{equation}\label{23}
  g(U)=\frac{U^2}{2\lambda_\beta}+N\ln(\bar{E}{\bf 1_4}-U)\,.
\end{equation}
The second equation (\ref{16}) has been obtained by an integration by parts.

Equation (\ref{16}) is still exact, but now the limit $N\to\infty$ is performed. Since
$\lambda_\beta$ is of ${\cal O}(N)$ (see equation~(\ref{19})), $U$ and $\bar{E}$ in
equation~(\ref{23}), too, must be of ${\cal O}(N)$. $E$, on the other hand, is of the
order of the mean level spacing and thus of ${\cal O}(1)$. Furthermore, $\phi$ is of
${\cal O}(\frac{1}{\sqrt{N}})$. Consequently $\Tr g(U)$ is of ${\cal O}(N)$, whereas all
other terms entering the integral (\ref{16}) are of ${\cal O}(1)$.

This suggests to diagonalize $U$,
\begin{equation}\label{21}
  U=T^{-1}U_DT\,,
\end{equation}
and perform the integrations over the elements of the diagonal matrix $U_D$ by means of
the saddle point technique. The saddle points are obtained from the zeros of $g'(u)$,
whence follows
\begin{equation}\label{24}
  u_{A/R}=\frac{N}{\pi}\left(\frac{\pi\bar{E}}{2 N}\pm\imath
  \sqrt{1-\left(\frac{\pi\bar{E}}{2 N}\right)^2}\right)\,.
\end{equation}
The plus and the minus sign belong to the advanced saddle point $u_A$, and the retarded
one $u_R$, respectively. The matrix $U_D$ at the saddle point is thus given by
\begin{equation}\label{25}
  \left(U_D\right)_S=\left(\begin{array}{cc}
    u_A{\bf 1_2} & 0 \\
    0 & u_R{\bf 1_2} \
  \end{array}\right)\,.
\end{equation}

The matrix $T$ diagonalizing $U$ may be parameterized as
\begin{equation}\label{21a}
  T=\left(\begin{array}{cc}
    \sqrt{1+\bet\gam} & \imath\bet \\
    -\imath\gam & \sqrt{1+\gam\bet} \
  \end{array}\right)\,,
\end{equation}
where $\bet$ and $\gam$ are $2\times 2$ supermatrices. Inserting equations (\ref{25}) and
(\ref{21a}) into equation [\ref{21}), we obtain for the matrix $U$ at the saddle point
\begin{equation}\label{26}
  U_S=\left(\begin{array}{cc}
    u_A{\bf 1_2}+\imath\Delta\bet\gam & \Delta\bet\sqrt{1+\gam\bet}\\
     \Delta\gam\sqrt{1+\bet\gam} & u_R{\bf 1_2}+\imath\Delta\gam\bet
    \end{array}\right)\,,
\end{equation}
where

\begin{equation}\label{26a}
  \Delta=\frac{2N}{\pi}\sqrt{1-\left(\frac{\pi\bar{E}}{2 N}\right)^2}=\frac{2N}{\pi}\rho\,,
\end{equation}
In the last equation we used expression (\ref{18}) for the mean density of states $\rho$.

 We are now left with
\begin{eqnarray}\label{22}
  \lefteqn{\ree}\nonumber\\
  &\sim&\frac{1}{N\lambda_\beta^2}\left<\Tr\left(U_{AR}\,\sigma\,
  U_{RA}\sigma\right)
 e^{-\frac{1}{2\lambda_\beta}\Tr
  \left[E\left(U_{AA}-U_{RR}\right)+\phi^2\left(U_{AR}U_{RA}\right)\right]}\right>\,,
  \nonumber\\
\end{eqnarray}
where only terms in $\phi$ surviving the $N\to\infty$ limit have been taken. The brackets
denote the average over the angular variables entering the matrix $T$, taken at the
saddle point. Using equation (\ref{26}) we obtain for the quantities entering on the
right hand side of equation (\ref{22})
\begin{eqnarray}\label{27}
  \Tr \left(U_{AA}-U_{RR}\right)&=&2\imath\Delta\Tr\bet\gam
  \,,\\\label{28}
  \Tr U_{AR}U_{RA}&=&\Delta^2\Tr\left[\bet\gam+(\bet\gam)^2\right]
  \,,\\\label{29}
  \Tr(U_{AR}\sigma U_{RA}\sigma)&=&\Delta^2\Tr(\gam\sqrt{1+\bet\gam}\sigma\bet\sqrt{1+\gam\bet}
  \sigma)
  \,.
\end{eqnarray}

The matrices $\bet$ and $\gam$ are diagonalized by means of the transformation
\begin{equation}\label{30}
  \bet=P\bet_DQ^{-1}\,,\quad \gam=Q\gam_DP^{-1}\,,
\end{equation}
where
\begin{equation}\label{31}
 \bet_D=\left(\begin{array}{cc}
    t & \cdot \\
    \cdot & \imath\bar{t} \
  \end{array}\right)\,,\qquad
\gam_D=\left(\begin{array}{cc}
    t^* & \cdot \\
    \cdot & \imath\bar{t}^* \
  \end{array}\right)\,,
\end{equation}
and
\begin{equation}\label{32}
 \fl P=\left(\begin{array}{cc}
    \sqrt{1+\alpha\alpha^*} & \alpha \\
    \alpha^* & \sqrt{1+\alpha^*\alpha} \
  \end{array}
\right)\,,\qquad
  Q=\left(\begin{array}{cc}
    \sqrt{1+\beta\beta^*} & \beta \\
    \beta^* & \sqrt{1+\beta^*\beta} \
  \end{array}\right)\,,
\end{equation}
(see e.\,g. Chapter 10 of reference \cite{haa01b}). Inserting these expressions into
equations (\ref{27}) to (\ref{29}), we obtain,
\begin{eqnarray}\label{33}
 \frac{E}{2\lambda_\beta}\Tr \left(U_{AA}-U_{RR}\right)&=&2\pi\imath\rho E
 \left(tt^*+\bar{t}\bar{t}^*\right)\,,\\
 \label{33a}
 \frac{\phi^2}{2\lambda_\beta}\Tr U_{AR}U_{RA}&=&\frac{\epsilon}{2}\rho^2\left[
  tt^*+\bar{t}\bar{t}^*+(tt^*)^2-(\bar{t}\bar{t}^*)^2
  \right]\,,\\
 \Tr(U_{AR}\sigma U_{RA}\sigma)&=&\Delta^2\Tr(\gam_D\sqrt{1+\bet_D\gam_D}\sigma_P
 \bet_D\sqrt{1+\gam_D\bet_D}\sigma_Q)\,,\nonumber\\
\end{eqnarray}
where expression (\ref{19}) for $\lambda_\beta$ and expression (\ref{00i}) for $\epsilon$
were used, and
\begin{eqnarray}\label{33b}
  \sigma_P&=&P^{-1}\sigma P=\left(\begin{array}{cc}
    1+2\alpha\alpha^* & 2\alpha \\
    -2\alpha^* & -1-2\alpha^*\alpha
  \end{array}\right)\,,\\\label{33c}
  \sigma_Q&=&Q^{-1}\sigma Q=\left(\begin{array}{cc}
    1+2\beta\beta^* & 2\beta \\
    -2\beta^* & -1-2\beta^*\beta
  \end{array}\right)\,.
\end{eqnarray}
For the calculation of the average (\ref{22}) over the angular variables the `surface
volume' element is needed,
\begin{equation}\label{34}
  d[\Omega]=\frac{dt\,dt^*\,d\bar{t}\,d\bar{t}^*\,d\alpha\,d\alpha^*\
  d\beta\,d\beta^*}{\left(tt^*+\bar{t}\bar{t}^*\right)^2}
\end{equation}
(see again reference \cite{haa01b}). The integral over the anticommuting variables is
easily performed. Only $\sigma_P$ and $\sigma_Q$ depend on the variables $\alpha,
\alpha^*$, and $\beta,\beta^*$, respectively, and the corresponding integrals reduce to
\begin{equation}\label{34a}
  \int d\alpha\,d\alpha^*\ \sigma_P\sim{\bf 1_2}\,,\quad
 \int d\beta\,d\beta^*\ \sigma_Q\sim{\bf 1_2}\,,
\end{equation}
whence follows
\begin{eqnarray}\label{35}
  \int d\alpha\,d\alpha^*\
  d\beta\,d\beta^* \Tr(U_{AR}\sigma U_{RA}\sigma)&\sim&
  \Tr\left(U_{AR}U_{RA}\right)\nonumber\\
  &\sim& \Delta^2\Tr\left[ \bet\gam+(\bet\gam)^2\right]\nonumber\\
  &\sim& \Delta^2\left[tt^*+\bar{t}\bar{t}^*+(tt^*)^2-(\bar{t}\bar{t}^*)^2
  \right]\,.
\end{eqnarray}
Collecting the results, we obtain from equation (\ref{22})
\begin{eqnarray}\label{36}
  \ree&\sim&\frac{1}{N}\left(\frac{\Delta}{\lambda_\beta}\right)^2
  \int dt\,dt^*\,d\bar{t}\,d\bar{t}^*
  \frac{tt^*+\bar{t}\bar{t}^*+(tt^*)^2-(\bar{t}\bar{t}^*)^2}
  {\left(tt^*+\bar{t}\bar{t}^*\right)^2}\nonumber\\&&\times
  e^{-2\pi\imath\rho E\left(tt^*+\bar{t}\bar{t}^*\right)}
  e^{-\frac{\epsilon}{2}\rho^2\left[tt^*+\bar{t}\bar{t}^*
  +(tt^*)^2-(\bar{t}\bar{t}^*)^2\right]}\,.
\end{eqnarray}

The $t,t^*$ integration is over the whole plane, whereas the $\bar{t},\bar{t}^*$
integration is restricted to the unit circle $\bar{t}\bar{t}^*\le 1$. Introducing polar
variables, we obtain
\begin{eqnarray}\label{37}
  \ree&\sim&\frac{\rho^2}{N}
  \int\limits_0^\infty dx\int\limits_0^1 dy
  \frac{x+y+x^2-y^2}
  {(x+y)^2}\nonumber\\&&\times
  e^{-2\pi\imath\rho E(x+y)}
  e^{-\frac{\epsilon}{2}\rho^2(x+y)(1+x-y)}\,.
\end{eqnarray}
Inserting this result into equation~(\ref{02}), and introducing $\bar{E}=(E_1+E_2)/2$ and
$E=E_1-E_2$ as new integration variables, we get, fixing the constant of proportionality
by the condition $f_\epsilon(0)=1$,
\begin{eqnarray}\label{38}
  f_\epsilon(\tau)&=&\frac{1}{N}\int d\bar{E}\rho^2
  \int\limits_0^\infty dx\int\limits_0^1 dy
  \frac{1+x-y}
  {x+y}\nonumber\\&&\times
  \delta [\tau-\rho(x+y)]
  e^{-\frac{\epsilon}{2}\rho^2(x+y)(1+x-y)}\,.
  \end{eqnarray}
The $\bar{E}$ integration is nothing but an energy average. Restricting the discussion to
the band centre, we may discard this average and obtain
\begin{equation}\label{40}
  f_\epsilon(\tau)=\frac{1}{\tau}\int_0^{{\rm Min}(\tau,1)}dy
  (1+\tau-2y)e^{-\frac{\epsilon}{2}\tau(1+\tau-2y)}\,.
\end{equation}
The integral is easily performed with the result
\begin{equation}\label{41}
  f_\epsilon(\tau)=\left\{\begin{array}{ll}
    e^{-\frac{\epsilon}{2}\tau}\left[s(\frac{\epsilon}{2}\tau^2)
    -\tau s'(\frac{\epsilon}{2}\tau^2)\right]\,,\qquad &\tau\le1 \\
    e^{-\frac{\epsilon}{2}\tau^2}\left[s(\epsilon\tau)
    -\frac{1}{\tau}s'(\frac{\epsilon}{2}\tau)\right]\,,&\tau>
    1
  \end{array}\right.\,,
\end{equation}
where
\begin{equation}\label{42}
  s(x)=\frac{\sinh(x)}{x}\, ,
\end{equation}
and $s'(x)$ denotes its derivative.

\begin{figure}
  \includegraphics[width=7cm]{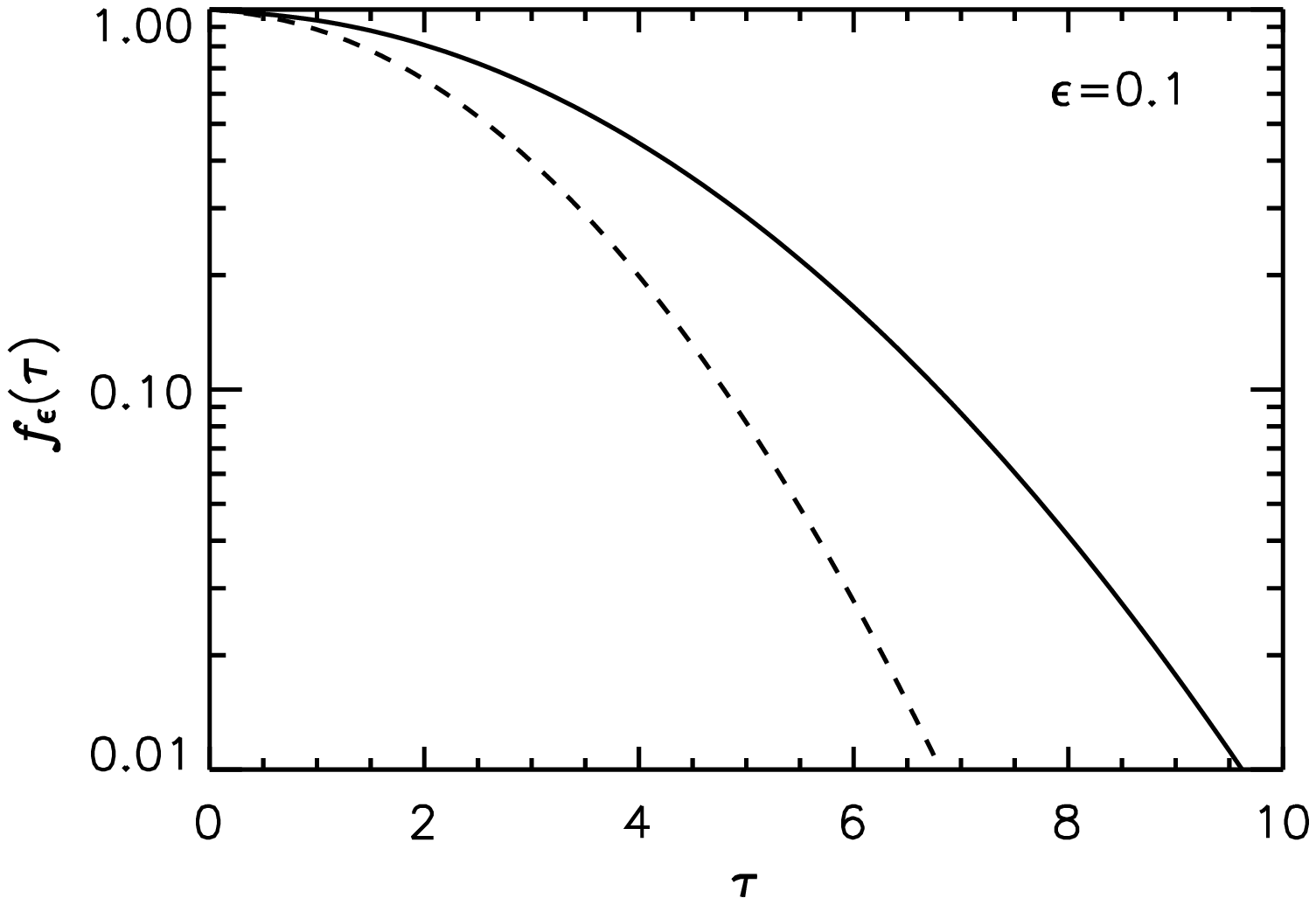}\hfill
  \includegraphics[width=7cm]{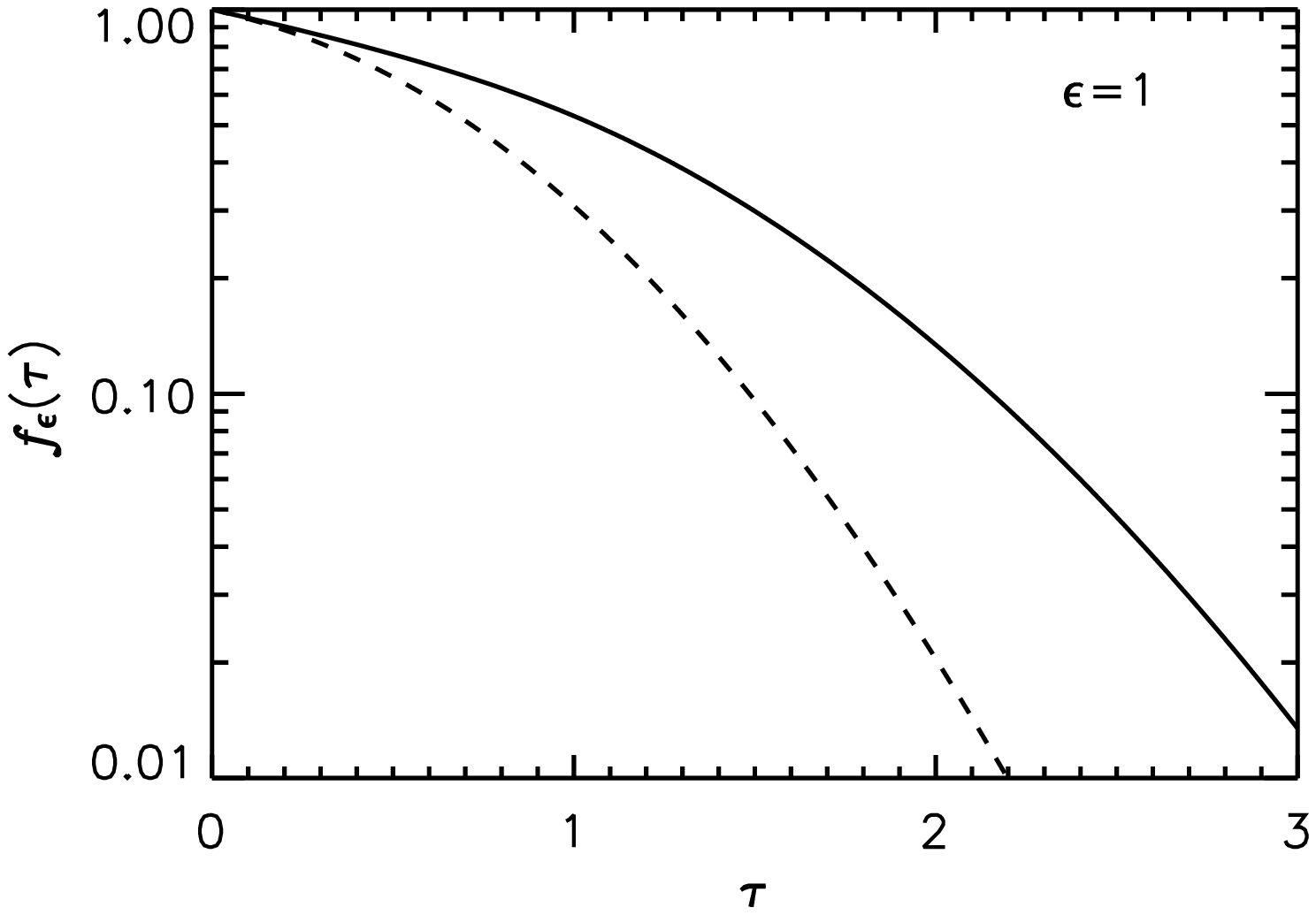}\\
  \includegraphics[width=7cm]{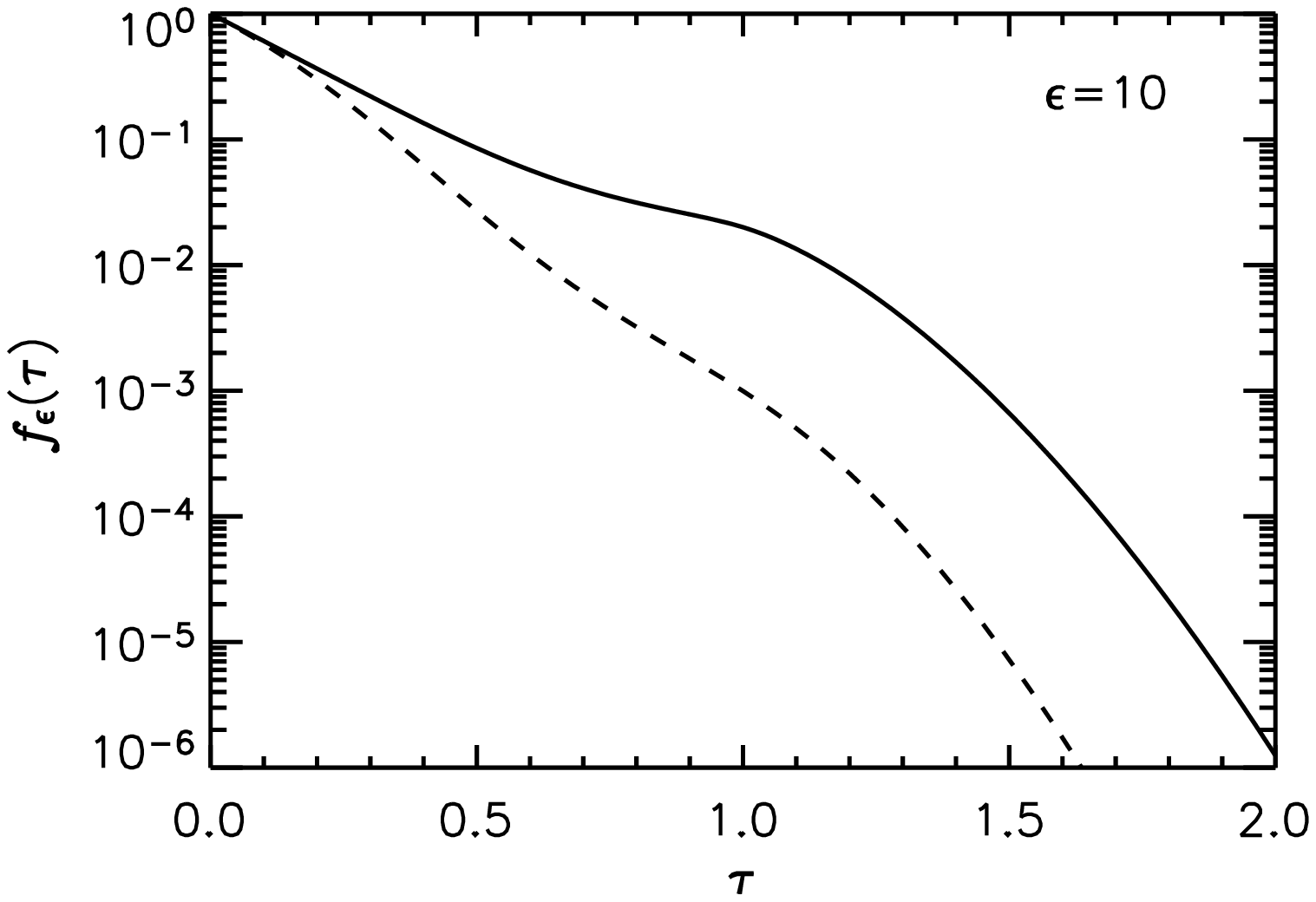}\hfill
  \includegraphics[width=7cm]{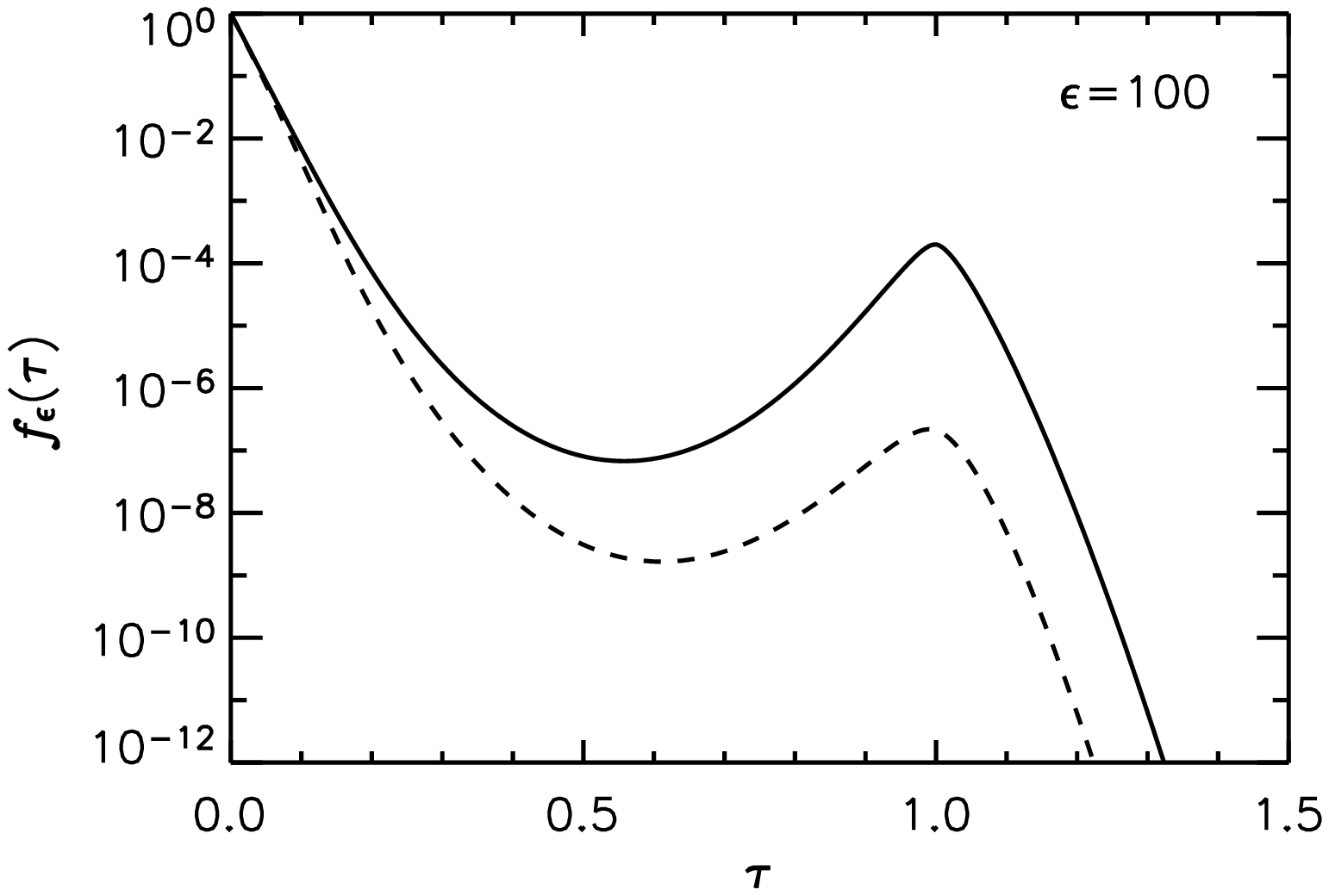}\\
  \caption{Fidelity amplitude $f_\epsilon(\tau)$ for the GUE (solid lines) and the GOE (dashed lines) for different values of the perturbation
  strength $\epsilon$. $\tau$ is given in units of the Heisenberg time.}\label{fig2}
\end{figure}

Equation (\ref{41}) is the central result of this section. It gives an analytic
expression for the GUE average of the fidelity amplitude for arbitrary  perturbation
strengths. $f_\epsilon(\tau)$ and its first derivative are continuous, but the second
derivative shows a discontinuity at $\tau=1$. A similar situation is known for the
spectral form factor, where, however, for the GUE already the first derivative is
discontinuous.

The solid lines in Figure \ref{fig2} show the GUE fidelity amplitude for different values
of the perturbation strength $\epsilon$. For $\epsilon\ll1$ the fidelity decay is
predominantly Gaussian. For $\epsilon=1$ and small times $\tau$ an exponential decay is
found, with a cross-over to Gaussian behaviour at $\tau=1$, both observations in
accordance with results known from literature. For $\epsilon\gg1$ the fidelity decay is
exponential for short times. Close to $\tau=1$, however, there is a conspicuous partial
revival of the fidelity which had not been reported before, as it seems. For still longer
times the decay becomes Gaussian again.

In the limit of small perturbations equation (\ref{41}) reduces to
\begin{equation}\label{43}
  f_\epsilon(\tau)=\left\{\begin{array}{ll}
    1-\frac{\epsilon}{2}\left(\tau+\frac{\tau^3}{3}\right)\,,\qquad &\tau\le1 \\
    1-\frac{\epsilon}{2}\left(\frac{1}{3}+\tau^2\right)\,,&\tau>1
  \end{array}\right.\,,
\end{equation}
This is in complete accordance with the results obtained by Gorin \etal \cite{gor04}.

\section{The GOE case}

The first steps in the calculation of the ensemble average of the fidelity amplitude for
the GOE  are  the same as for the GUE. Equation (\ref{07a}) for $S$ remains correct, but
now $S$ is an $8\times 8$ matrix with $\zn$ given by
\begin{equation}\label{51}
  \zn=\frac{1}{\sqrt{2}}\left(\begin{array}{c}
    x_n \\ x_n^* \\ \xi_n \\ \xi_n^* \\
    y_n \\ y_n^* \\ \eta_n \\ \eta_n^*\
  \end{array}\right)\,,\qquad {\bf L}=\left(\begin{array}{cc}
    {\bf 1_4} & \cdot \\
    \cdot & -{\bf 1_4}\
  \end{array}\right)\,.
\end{equation}

In taking the adjoint of $\zn$ one has to consider that the complex conjugate of the
complex conjugate of an antisymmetric variable is defined as
$\left(\alpha^*\right)^*=-\alpha$, see the appendix of reference \cite{ver85a}.

Up to equation (\ref{23}) the further procedure is nothing but a step-by-step repetition
of the calculation for the GUE case. The main problem for the GOE case arises from the
diagonalization of matrix $U$,
\begin{equation}\label{53}
  U=T^{-1}U_DT\,,
\end{equation}
see equation (\ref{21}). Not all of the matrix elements of $S$ are different, as is
evident from its definition, with the consequence that $S$ obeys a number of symmetries
which are inherited by the matrix $U$. The matrices $T$ have to be chosen such that all
symmetries are conserved. It is a highly non-trivial task to find the best
parameterization for the matrix elements of $T$ obeying these constraints. Fortunately
this problem has already been solved by Verbaarschot, Weidenm\"uller, and Zirnbauer in
their disseminating work \cite{ver85a}. We just cite their essential results:

Equations (\ref{30}) still hold, but now $\bet_D$ and $\gam_D$ are equal and given by
\begin{equation}\label{54}
  \bet_D=\gam_D=\left(\begin{array}{cccc}
    t_1 & \cdot & \cdot & \cdot \\
    \cdot & t_2 & \cdot & \cdot \\
    \cdot & \cdot & \imath\bar{t} & \cdot \\
    \cdot & \cdot & \cdot & \imath\bar{t}
  \end{array}\right)\,.
\end{equation}

The parameterization of the matrices $P$ and $Q$ is complicated, and is given in the
appendices of reference \cite{ver85a}. For the present purpose it is sufficient to note
that the angular averages over the matrices $\sigma_P=P\sigma P^{-1}$ and
$\sigma_Q=Q\sigma Q^{-1}$ (see equations (\ref{33b}) and(\ref{33c})) up to a constant
factor again yield the unit matrix, as can be shown by explicit calculation. All formulas
of section \ref{gue} can thus be applied directly to the GOE situation. The surface
volume element for the only remaining variables $t_1, t_2, \bar{t}$ is given by
\begin{equation}\label{55}
  d[\Omega]=\frac{(1-\bar{t}^2)\bar{t}^3\left|t_1^2-t_2^2\right|}
  {\sqrt{(1+t_1^2)(1+t_2^2)}(t_1^2+\bar{t}^2)^2(t_2^2+\bar{t}^2)^2}
  dt_1\,dt_2\,d\bar{t} \,,
\end{equation}
and the integrations are from 0 to $\infty$ for $t_1$ and $t_2$, and from 0 to 1 for
$\bar{t}$.

Collecting the results, and proceeding in exactly the same way as for the GOE case, we
finally end up with
\begin{eqnarray}\label{56}
  f_\epsilon(\tau)&\sim&\int\limits_0^\infty dx\int\limits_0^\infty dy \int\limits_0^1 dz\,
  \frac{(1-z)z|x-y|}{\sqrt{x(1+x)y(1+y)}(x+z)^2(y+z)^2}\nonumber\\
  &&\times [x(1+x)+y(1+y)+2z(1-z)]\nonumber\\
  &&\times e^{-\frac{\epsilon}{2}[x(1+x)+y(1+y)+2z(1-z)]}\,
  \delta\left[(x+y)/2+z-\tau\right]\,.
\end{eqnarray}

Substituting $u=(x+y)/2$ and $v=(x-y)/2$, we obtain
\begin{eqnarray}\label{57}
  f_\epsilon(\tau)&=&2\int\limits_{{\rm Max}(0,\tau-1)}^\tau du\int\limits_0^u
  \frac{v\,dv}{\sqrt{[u^2-v^2][(u+1)^2-v^2]}}
  \frac{(\tau-u)(1-\tau+u)}{(v^2-\tau^2)^2}\nonumber\\
  &&\times [(2u+1)\tau-\tau^2+v^2]e^{-\frac{\epsilon}{2}[(2u+1)\tau-\tau^2+v^2]}\,,
\end{eqnarray}
where the constant of proportionality again was fixed by the condition $f_\epsilon(0)=1$.
For $\epsilon=0$ the right hand side of equation (\ref{57}) must be one by construction,
but it is not straightforward to show this explicitly. Since the corresponding
calculation may be of some interest, it is reproduced in \ref{app1}.

Equation (\ref{57}) gives an explicit expression for the fidelity amplitude for the GOE
case. It is not yet suited directly for a numerical integration, since the integrand
contains a number of singularities. But it is not difficult to remove them by suitable
substitutions of integration variables. This is done in \ref{app2}.

The dashed lines in Figure \ref{fig2} show the results of the calculation for the same
$\epsilon$ parameters as before. We notice that the partial recovery of the fidelity
close to $\tau=1$ is still present for large $\epsilon$ values, but is considerably less
pronounced than for the GUE case.

\section{Numerical simulations}

In this section we present random matrix simulations to affirm the analytical findings
for the Gaussian orthogonal and unitary ensembles. Further we show numerical results for
the Gaussian symplectic ensemble which has not been treated analytically.

In our simulations the Hamiltonians $H_0$ and $H_1$ are random matrices of dimension $N
\times N$ with variances of the diagonal and off-diagonal elements given by equation
(\ref{18a}). To calculate the fidelity amplitude, we write expression (\ref{01}) as
\begin{eqnarray}
\label{58}
  f_\epsilon(\tau) &=& \frac{1}{N}\left<{\rm Tr}\left[R_\phi \rme^{2\pi\imath H_\phi^D \tau} R_\phi^{-1}
 R_0 \rme^{-2\pi\imath H_0^D \tau} R_0^{-1} \right]\right> \nonumber\\
 &=&  \frac{1}{N}\left<{\rm Tr}\left[ \rme^{2\pi\imath H_\phi^D \tau} R
  \rme^{-2\pi\imath H_0^D \tau} R^{-1} \right]\right> \nonumber\\
 &=&  \frac{1}{N} \left< \sum\limits_{kl} \rme^{2\pi\imath\tau(E_k^{(\phi)}-E_l^{(0)})} |R_{lk}|^2 \right> \, ,
\end{eqnarray}
where $H_0^D=R_0^{-1}H_0 R_0$ and $H_\phi^D=R_\phi^{-1}H_\phi R_\phi$ are diagonal, and
$R=R_\phi^{-1}R_0$.

In the numerical simulations the trace in equation (\ref{58}) was restricted to 20
percent of the eigenvalues in the centre of the spectrum where the mean level density is
still about constant.  The average was taken over up to 8000 random matrices for $H_0$,
and for each of them over 50 random matrices for $H_1$. For larger values of the
perturbation strength $\epsilon$ it became more and more important to choose the
dimension $N$ of the matrices large enough to avoid finite-size effects. $N=500$ proved to be sufficient for $\epsilon\le10$.

\begin{figure}
\begin{center}
  \includegraphics[width=0.6\textwidth]{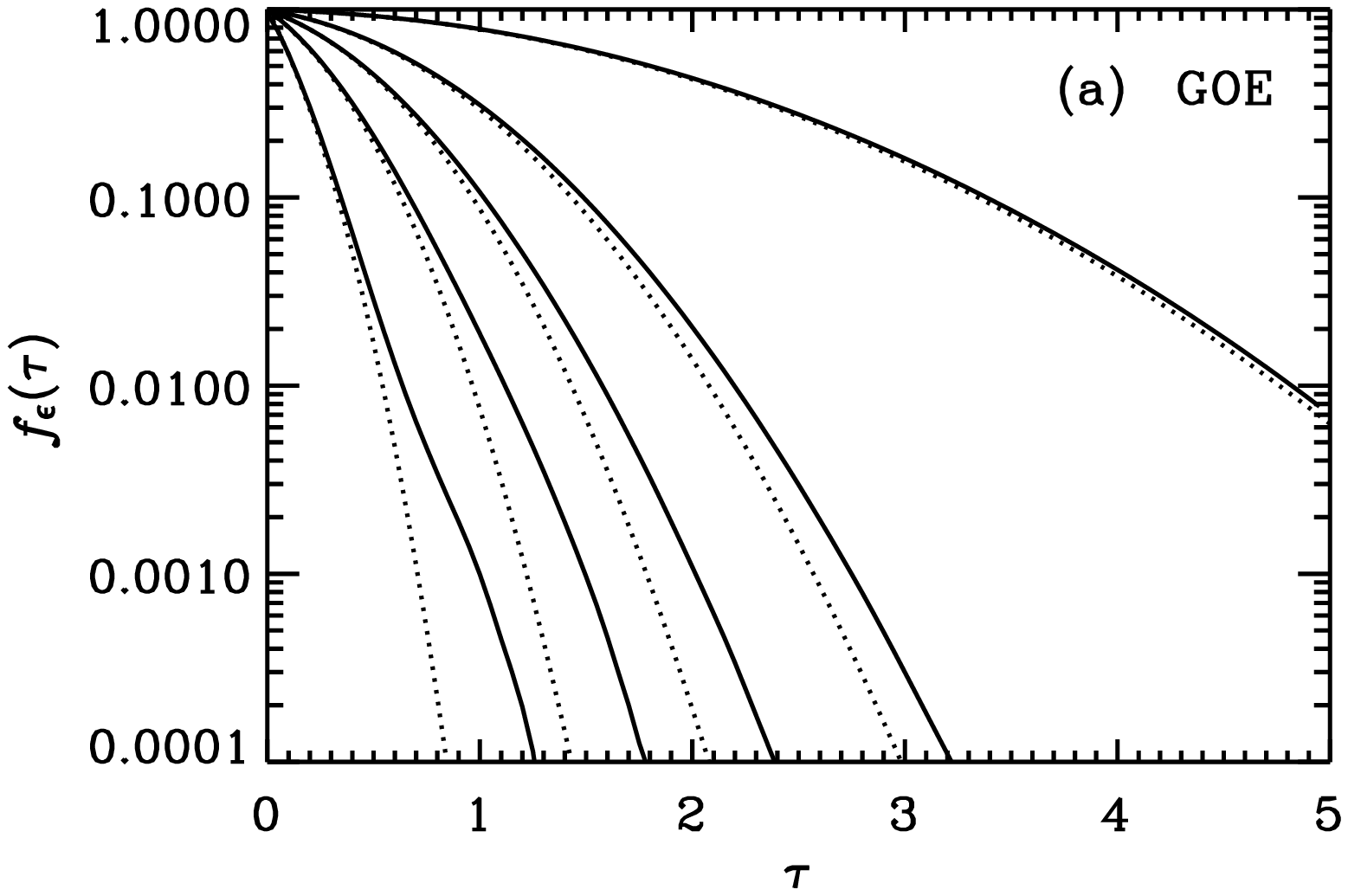}\\
  \includegraphics[width=0.6\textwidth]{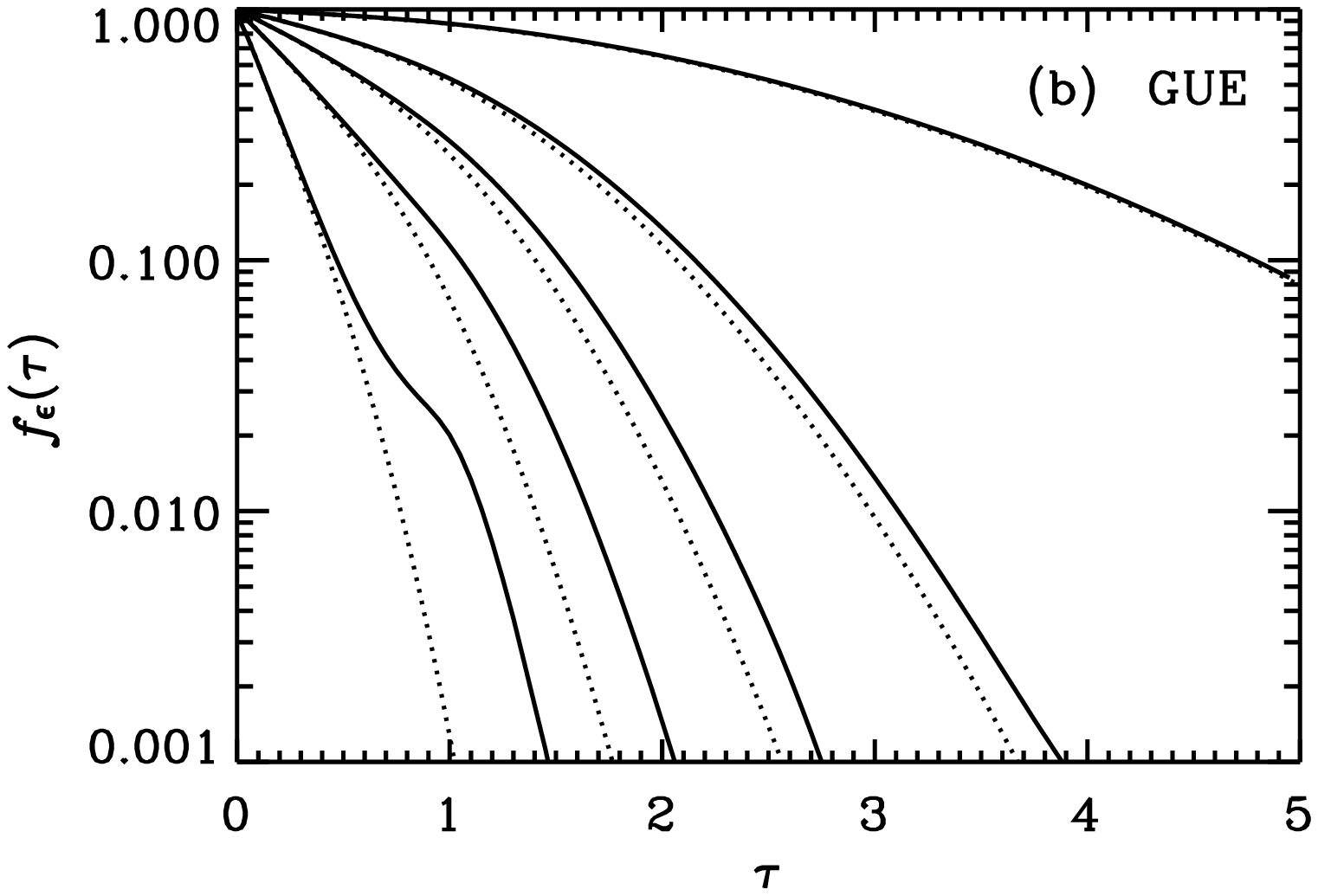}\\
  \includegraphics[width=0.6\textwidth]{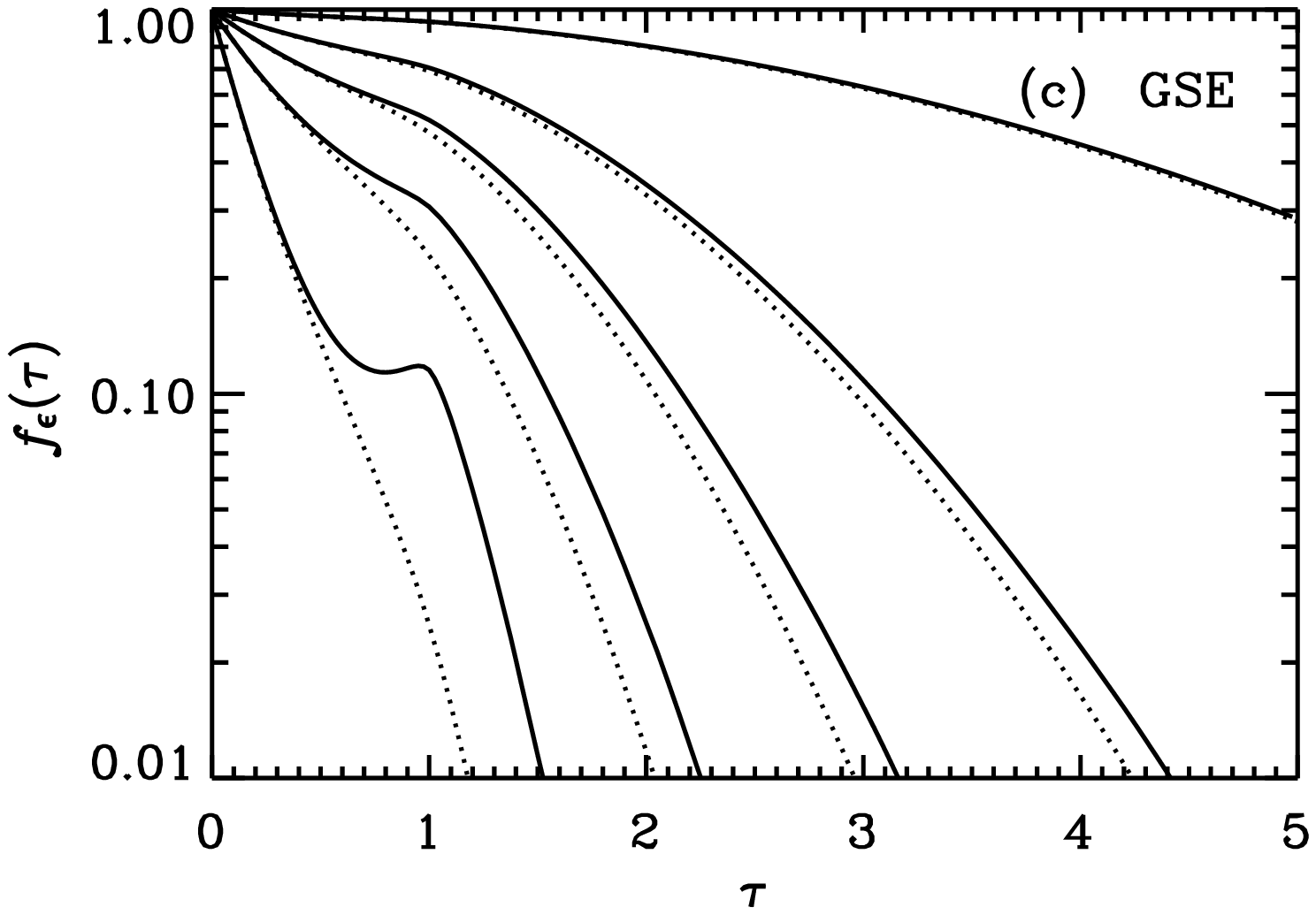}
  \caption{Fidelity amplitude $f_\epsilon(\tau)$ for the GOE (a), the GUE (b),
  and the GSE (c)  for $\epsilon=$ 0.2, 1, 2, 4 and 10.
   The solid lines show the results of the numerical simulations, and the dotted lines
   those of the linear response approximation.  For the GOE and the GUE
   the numerical results are in agreement with the analytical
   results within the limits of the line strength.
    }\label{fig:vgl_gue_goe}
\end{center}
\end{figure}

The results for the three Gaussian ensembles are shown in Figure \ref{fig:vgl_gue_goe}.
For the GOE and the GUE the numerical simulations are in perfect agreement with the
analytical result for all $\epsilon$ values shown. For comparison,  the fidelity
amplitudes in the linear response approximation \cite{gor04} (see equations (\ref{00a})
and (\ref{00aa})) are shown as well. For small perturbation strengths and small values of
$\tau$, the linear response result is a good approximation, but the limits of its
validity are also clearly illustrated. In particular, it does not show any indication of
the recovery near $\tau=1$.

\section{Discussion}

This work extends the results by Gorin \etal \cite{gor04} to the regime of strong perturbations
using supersymmetry techniques.
An intuitive explanation for the surprising recovery of the fidelity amplitude at the Heisenberg time can be given in terms of the Brownian-motion model for the eigenvalues of random matrices. The behaviour of the fidelity amplitude has its direct
analogue in the Debye-Waller factor of solid state physics (see reference \cite{stoe1}).
It is stressed that our result is generic and not restricted to random matrix systems. 
For instance, the fidelity recovery has recently been observed in a spin-chain model by Pineda \etal \cite{pin}.

The results of the present work may be easily extended to all situations, where the
Gaussian averages (see equation (\ref{04})) lead to expressions allowing a subsequent
Hubbard-Stratonovich transformation. This is, e.\,g., the case, if $H_0$ is taken from
the GOE, and $H_1$ is from the GUE, or is purely imaginary antisymmetric. Perturbations,
where the diagonal is zero, are of particular interest, since in such a situation the
decay of fidelity freezes \cite{pro03b,pro}.

There might be still another application of the formulas derived in this paper. If
$f_\epsilon(\tau)$ is expanded into a power series of $\epsilon$, the linear term can be
expressed in terms of the spectral form factor, i.\,e. the Fourier transform of the
two-point correlation function \cite{gor04}. In a similar way the coefficients of the
$n$th power of $\epsilon$ depend on all $k$-point correlation functions up to $k=n$.
$f_\epsilon (\tau)$ may thus be used as a generating function to obtain these terms in a
simple way.

\ack

This paper has profited a lot from numerous discussions on the subject of fidelity with
Thomas Seligman, Cuernavaca, Mexico, Thomas Gorin, Freiburg, Germany and Toma{\v z}
Prosen, Ljubljana, Slovenia. Thomas Guhr, Lund, Sweden is thanked for discussions of the
supersymmetry aspects of this paper. The work was supported by the Deutsche
Forschungsgemeinschaft.

\appendix

\section{Proof of $f_0(\tau)\equiv 1$}\label{app1}

Substituting $x=\sqrt{u^2-v^2}$, we obtain from equation (\ref{57})
\begin{eqnarray}\label{a01}
  f_\epsilon(\tau)&=&2\int\limits_{{\rm Max}(0,\tau-1)}^\tau du\int\limits_0^u
  \frac{dx}{\sqrt{x^2+2u+1}}
  \frac{(\tau-u)(1-\tau+u)}{(\tau^2-u^2+x^2)^2}\nonumber\\
  &&\times [(2u+1)\tau-\tau^2+u^2-x^2]e^{-\frac{\epsilon}{2}[(2u+1)\tau-\tau^2+u^2-x^2]}\,.
\end{eqnarray}
Specializing to $\epsilon=0$, and applying the substitution
\begin{equation*}
  x=(2u+1)\left[ \frac{1}{2}\left(\sqrt{2z+1}+\frac{1}{\sqrt{2z+1}}\right)\right]
\end{equation*}
this may be written as
\begin{eqnarray}\label{a02}
  f_0(\tau)&=&2\int\limits_{{\rm Max}(0,\tau-1)}^\tau du\,(\tau-u)(1-\tau+u)\nonumber\\
  &&\times\left[(u^2+2\tau u +\tau-\tau^2)I_2-(2u+1)I_3\right]\,,
\end{eqnarray}
where
\begin{eqnarray}\label{a03}
  I_2&=&\int\limits_0^u dz
  \frac{2z+1}{\left[(\tau^2-u^2)(2z+1)+(2u+1)z^2\right]^2}\,,\\
   I_3&=&\int\limits_0^u dz \frac{z^2}{\left[(\tau^2-u^2)(2z+1)+(2u+1)z^2\right]^2}\,.
\end{eqnarray}

The same equations can be found already in reference \cite{gor02a}. The latter two
integrals can be performed elementary and yield
\begin{eqnarray}\label{a04}
  I_2=\frac{q^2-p^2}{2p^3q^3}\arctan\frac{uq}{(u+1)p}+\frac{u(u+1)}{2\tau^2p^2q^2}\,,\\
  I_3=\frac{1}{2pq^3}\arctan\frac{uq}{(u+1)p}-\frac{u(u+1)}{2\tau^2(2u+1)q^2}\,,
\end{eqnarray}
where
\begin{equation}\label{a05}
  p=\sqrt{\tau^2-u^2}\,,\qquad q=\sqrt{(u+1)^2-\tau^2}\,.
\end{equation}

Let us denote the part of $f_0(\tau)$, depending in the integrand of the arctan terms, by
$f_0^a(\tau)$. It may be transformed as
\begin{eqnarray}\label{a06}
 \fl f_0^a(\tau)
  &=& \int\limits_{{\rm Max}(0,\tau-1)}^\tau du
  \left[(u^2+2\tau u+\tau-\tau^2)\left(\frac{1}{p^2}-\frac{1}{q^2}\right)
  -(2u+1)\frac{1}{q^2}\right]\nonumber\\
  &&\times (\tau-u)(1+u-\tau)\frac{1}{pq}\arctan\frac{uq}{(u+1)p}\nonumber\\
  &=& \int\limits_{{\rm Max}(0,\tau-1)}^\tau du
  \left[\frac{u^2+2\tau u+\tau-\tau^2}{(\tau-u)(\tau+u)}
  -\frac{u^2+2u(\tau+1)+\tau-\tau^2+1}{(u+1-\tau)(u+1+\tau)}\right]\nonumber\\
  &&\times\sqrt{\frac{(\tau-u)(u+1-\tau)}{(\tau+u)(u+1+\tau)}}\arctan\frac{uq}{(u+1)p}
  \nonumber\\
  &=& -\int\limits_{{\rm Max}(0,\tau-1)}^\tau du
  \left[2+\frac{2u+1}{2}\left(\frac{1}{u-\tau}-\frac{1}{u+\tau}+\frac{1}{u+1-\tau}
  -\frac{1}{u+1+\tau}\right)\right]\nonumber\\
  &&\times\sqrt{\frac{(\tau-u)(u+1-\tau)}{(\tau+u)(u+1+\tau)}}\arctan\frac{uq}{(u+1)p}
  \nonumber\\
  &=& -\int\limits_{{\rm Max}(0,\tau-1)}^\tau du
  \left\{\left[2+(2u+1)\frac{\partial}{\partial u}\right]
  \sqrt{\frac{(\tau-u)(u+1-\tau)}{(\tau+u)(u+1+\tau)}}\right\}\arctan\frac{uq}{(u+1)p}
  \nonumber\\
&=& -\int\limits_{{\rm Max}(0,\tau-1)}^\tau du
  \left[2-\left\{\frac{\partial}{\partial u}(2u+1)\right\}\right]
  \sqrt{\frac{(\tau-u)(u+1-\tau)}{(\tau+u)(u+1+\tau)}}\arctan\frac{uq}{(u+1)p}\nonumber\\
  &&+\int\limits_{{\rm Max}(0,\tau-1)}^\tau du (2u+1)
  \sqrt{\frac{(\tau-u)(u+1-\tau)}{(\tau+u)(u+1+\tau)}}
  \frac{\partial}{\partial u}\arctan\frac{uq}{(u+1)p}\,,
  \nonumber\\
\end{eqnarray}
where in the last step an integration by parts was performed. The terms in the first row
cancel, and only the term in the last row survives. Performing the differentiation, we
have
\begin{equation}\label{a07}
  f_0^a(\tau)  = \int\limits_{{\rm Max}(0,\tau-1)}^\tau du
  \frac{1+3u(u+1)-\tau^2}{(u+\tau)(u+1+\tau)}\,.
\end{equation}
Collecting the results we are left with
\begin{eqnarray}\label{a08}
  f_0(\tau)&=& \int\limits_{{\rm Max}(0,\tau-1)}^\tau du
  \frac{1}{(u+\tau)(u+1+\tau)}\Bigg[(u^2+2\tau u+\tau-\tau^2)\frac{u(u+1)}{\tau^2}
  \nonumber\\
  &&+(2u+1)\frac{u(u+1)(\tau^2-u^2)}{(2u+1)\tau^2}+1+3u(u+1)-\tau^2\Bigg]\nonumber\\
  &=&\frac{1}{\tau}\int\limits_{{\rm Max}(0,\tau-1)}^\tau du
  \frac{2u^3+3u^2\tau-\tau^3+3u^2+3u\tau+u+\tau}{(u+\tau)(u+1+\tau)}\nonumber\\
  &=&\frac{1}{\tau}\int\limits_{{\rm Max}(0,\tau-1)}^\tau du (2u-\tau+1)\nonumber\\
  &=&\left.\frac{1}{\tau}u(u-\tau+1)\right|_{{\rm Max}(0,\tau-1)}^\tau\nonumber\\
  &=& 1\,,
\end{eqnarray}

q.\,e.\,d.

\section{Transformation of the integral (\ref{57})}\label{app2}

To turn equation (\ref{57}) into a form being suited for a numerical calculation, we
start with equation (\ref{a01}) by substituting $u=\tau\sin\phi$ and obtain
\begin{eqnarray}\label{b01}
  f_\epsilon(\tau)&=&2\int\limits_{{\rm Max}\left(0,\arcsin[(\tau-1)/\tau]\right)}^{\pi/2}
  d\phi\,\tau\cos\phi\,\tau(1-\sin\phi)[1-\tau(1-\sin\phi)]\nonumber\\
  &&\times \int\limits_0^{\tau\sin\phi}\frac{dx}{\sqrt{x^2+2\tau\sin\phi+1}
  \left(x^2+\tau^2\cos^2\phi\right)^2}Ze^{-\frac{\epsilon}{2} Z}\,,
\end{eqnarray}
where
\begin{equation}\label{b02}
  Z=(2\tau\sin\phi+1)\tau-\tau^2\cos^2\phi-x^2\,.
\end{equation}
Next we substitute $x=\hat{x}\tau\cos\phi$ with the result
\begin{eqnarray}\label{b03}
  f_\epsilon(\tau)&=&2\int\limits_{{\rm Max}\left(0,\arcsin[(\tau-1)/\tau]\right)}^{\pi/2}
  d\phi\,\frac{1-\tau(1-\sin\phi)}{1+\sin\phi}\nonumber\\
  &&\times \int\limits_0^{\tan\phi}\frac{d\hat{x}}
  {\sqrt{\hat{x}^2\tau^2\cos^2\phi+2\tau\sin\phi+1}
  \left(1+\hat{x}^2\right)^2}\hat{Z}e^{-\frac{\epsilon}{2}\tau \hat{Z}}\,,
\end{eqnarray}
where
\begin{equation}\label{b04}
  \hat{Z}=2\tau\sin\phi+1-\tau\cos^2\phi(1+\hat{x}^2)\,.
\end{equation}
After the final substitution $\hat{x}=\tan\alpha$ we end with
\begin{eqnarray}\label{b05}
  f_\epsilon(\tau)&=&2\int\limits_{{\rm Max}\left(0,\arcsin[(\tau-1)/\tau]\right)}^{\pi/2}
  d\phi\,\frac{1-\tau(1-\sin\phi)}{1+\sin\phi}\nonumber\\
  &&\times \int\limits_0^{\phi}d\alpha
  \frac{\cos\alpha\left[(2\tau\sin\phi+1)\cos^2\alpha-\tau\cos^2\phi\right]}
  {\sqrt{\tau^2\cos^2\phi\sin^2\alpha+(2\tau\sin\phi+1)\cos^2\alpha}}
  \,e^{-\frac{\epsilon}{2}\tau \hat{Z}}\,,
\end{eqnarray}
where now

\begin{equation}\label{b06}
  \hat{Z}=2\tau\sin\phi+1-\tau\frac{\cos^2\phi}{\cos^2\alpha}\,.
\end{equation}

The integrand of the double integral (\ref{b05}) behaves well everywhere, and the
numerical integration does not pose problems any longer.

\section*{References}


\begin{thebibliography}{10}

\bibitem{per84}
Peres A
\newblock 1984 {\em Phys. Rev. A} {\bf 30}~1610

\bibitem{cer02}
Cerruti N R and Tomsovic S
\newblock 2002 {\em Phys. Rev. Lett.} {\bf 88}~054103

\bibitem{jac01b}
Jacquod P, Silvestrov P G  and Beenakker C W J
\newblock 2001 {\em Phys. Rev. E} {\bf 64}~055203

\bibitem{Jal01}
Jalabert R A and Pastawski H M
\newblock 2001 {\em Phys. Rev. Lett.} {\bf 86}~2490

\bibitem{pas95}
Pastawski H M, Levstein P R  and Usaj G
\newblock 1995 {\em Phys. Rev. Lett.} {\bf 75}~4310

\bibitem{zha92}
Zhang S, Meier B H  and Ernst R R
\newblock 1992 {\em Phys. Rev. Lett.} {\bf 69}~2149

\bibitem{gor04}
Gorin T, Prosen T  and Seligman T H
\newblock 2004 {\em New J. of Physics} {\bf 6}~20

\bibitem{meh91}
Mehta M L
\newblock 1991 {\em Random Matrices. 2nd edition}
\newblock  (San Diego: Academic Press)

\bibitem{ver85a}
Verbaarschot J J M, Weidenm\"uller H A  and Zirnbauer M R
\newblock 1985 {\em Phys. Rep.} {\bf 129}~367

\bibitem{haa01b}
Haake F
\newblock 2001 {\em Quantum Signatures of Chaos. 2nd edition}
\newblock  (Berlin: Springer)

\bibitem{stoe1}
St\"ockmann H J and Sch\"afer R
\newblock 2004 {\em Preprint} nlin.CD/0409021

\bibitem{pin}
Pineda C, Sch\"afer R, Prosen T  and Seligman T H
\newblock To be published

\bibitem{pro03b}
Prosen T, Seligman T H  and \v{Z}nidari\v{c} M
\newblock 2003 {\em Prog. Theor. Phys. Suppl.} {\bf 150}~200

\bibitem{pro}
Prosen T and \v{Z}nidari\v{c} M
\newblock 2004 {\em Preprint} quant-ph/0401142

\bibitem{gor02a}
Gorin T and Seligman T H
\newblock 2002 {\em Phys. Rev. E} {\bf 65}~026214

\end{thebibliography}

\end{document}